\documentclass{article}

\usepackage{amsmath}
\usepackage{amsfonts}%
\usepackage{amssymb}%
\usepackage{adjustbox}%
\usepackage{graphicx}%
\usepackage{graphics,color}%
\usepackage{natbib}%
\usepackage{subfig}
\usepackage[outdir=./]{epstopdf}
\usepackage[plain,noend]{algorithm2e}
\newtheorem{theorem}{Theorem}

\newenvironment{proof}[1][Proof]{\textbf{#1.} }{\ \rule{0.5em}{0.5em}}

\usepackage{times}
\usepackage{bm}

\usepackage[plain,noend]{algorithm2e}

\makeatletter
\renewcommand{\algocf@captiontext}[2]{ \quad #1\algocf@typo. \AlCapFnt{}#2} % text of caption
\def\@algocf@capt@plain{top}
\renewcommand{\algocf@makecaption}[2]{%
  \addtolength{\hsize}{\algomargin}%
  \sbox\@tempboxa{\algocf@captiontext{#1}{#2}}%
  \ifdim\wd\@tempboxa >\hsize%     % if caption is longer than a line
    \hskip .5\algomargin%
    \parbox[t]{\hsize}{\algocf@captiontext{#1}{#2}}% then caption is not centered
  \else%
    \global\@minipagefalse%
    \hbox to\hsize{\box\@tempboxa}% else caption is centered
  \fi%
  \addtolength{\hsize}{-\algomargin}%
}
\makeatother

\def \A {\boldsymbol{A}}
\def \B {\boldsymbol{B}}

\def \Z {\boldsymbol{Z}}
\def \O {\boldsymbol{O}}
\def \R {\boldsymbol{R}}
\def \U {\boldsymbol{U}}
\def \V {\boldsymbol{V}}
\def \I {\boldsymbol{I}}
\usepackage{authblk}

\title{Optimal Prior Pooling from Expert Opinions}

\author[1]{A. KUME}
\affil[1]{School of Mathematics, Statistics and Actuarial Science, University of Kent,\\ Canterbury, U.K.
\textit{a.kume@kent.ac.uk}}

\author[2]{C. VILLA}
\affil[2]{School of Mathematics, Statistics and Physics, Newcastle University,\\ Newcastle Upon Tyne, U.K.
\textit{cristiano.villa@ncl.ac.uk}}

\author[3]{S. G. WALKER}
\affil[3]{Department of Statistics and Data Sciences, University of Texas at Austin, Austin, U.S.A. \textit{s.g.walker@math.utexas.edu}}

\begin{document}
\maketitle

\begin{abstract}
The pooling of prior opinions is an important area of research and has been for a number of decades. The idea is to obtain a single belief probability distribution from a set of expert opinion belief distributions.  The paper proposes a new way to provide a resultant prior opinion based on a minimization of information principle. This is done in the square-root density space, which is identified with the positive orthant of Hilbert unit sphere of differentiable functions.  It can be shown that the optimal prior is easily identified  as an extrinsic mean in the sphere. For distributions belonging to the exponential family, the necessary calculations are exact, and so can be directly applied. The idea can also be  adopted for any neighbourhood of a chosen base prior and spanned by a finite set of ``contaminating" directions.
\end{abstract}

%\begin{keywords}

\section{Introduction}

Among many models which deal with representation of continuous distributions, the recently explored Hilbert sphere of densities in \cite{KB15}, is one of the most elegant and useful in terms of ability to perform concrete calculations and derive simple geometric interpretations of results. 
The representation is motivated by the Fisher-Rao distance, which is the Riemannian geodesic  distance $\rho$ on the space of square-rooted densities $\psi=\sqrt{f}$ such that:  
$$\cos\rho(\psi_i,\psi_j)=<\psi_i,\psi_j>=\int \psi_i(x)\psi_j(x) dx,$$
for any pair $\psi_i, \psi_j \in \mathcal{X}$ where $\mathcal{X}=\left\{\psi|\psi(x)=\sqrt{f(x)}, x\in I \!\!R \text{ and  }f\in Dens\right\}$
and $Dens=\{f|\int_{ I \!\!R} f(x) dx=1\}$. 
It is easily seen that $\mathcal{X}$ is the intersection of the positive orthant with the Hilbert Unit sphere of dimension $S^{\infty}$, and so the corresponding spherical geometry is adopted for performing Bayesian analysis. 
In \cite{KB15}, the authors make use of this Riemannian manifold structure to explore the geodesic curves among the densities and use tangent spaces for linearizing the space locally via the exponential mapping.  All of these tools are available due to the spherical geometry, while are not generally available on alternative  geometric models, such as that of the \cite{Am85}, which is based on the parametric information metric. 

\begin{figure}[ht]
	\begin{minipage}[b]{0.45\linewidth}
\centering
		%\begin{figure}
		\includegraphics[scale=.1]{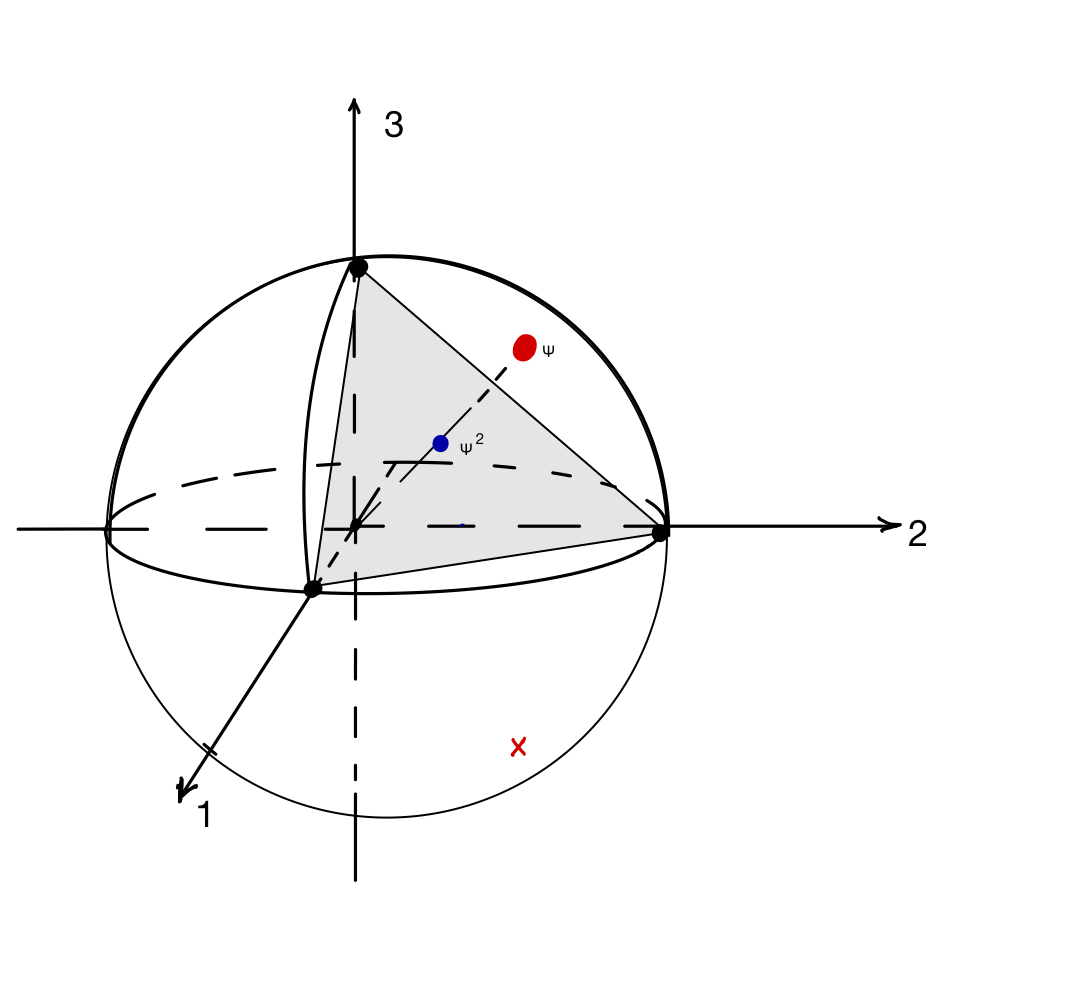} 
%	\caption{A representation of  the space of discrete measures of three outcomes. The component $\psi$ in red is a point on $S^2$, the sphere of dimension two. $\psi^2$ in blue is a point in the shaded simplex whose coordinates are squares of those of $\psi$,  so that the sum of coordinates equals  one. The red cross represents some reflection of $\psi$ outside the positive orthant.}
		%\end{figure}
	\end{minipage}
	\hspace{0.01cm}
	\begin{minipage}[b]{0.45\linewidth}
		\centering
		\includegraphics[scale=.1]{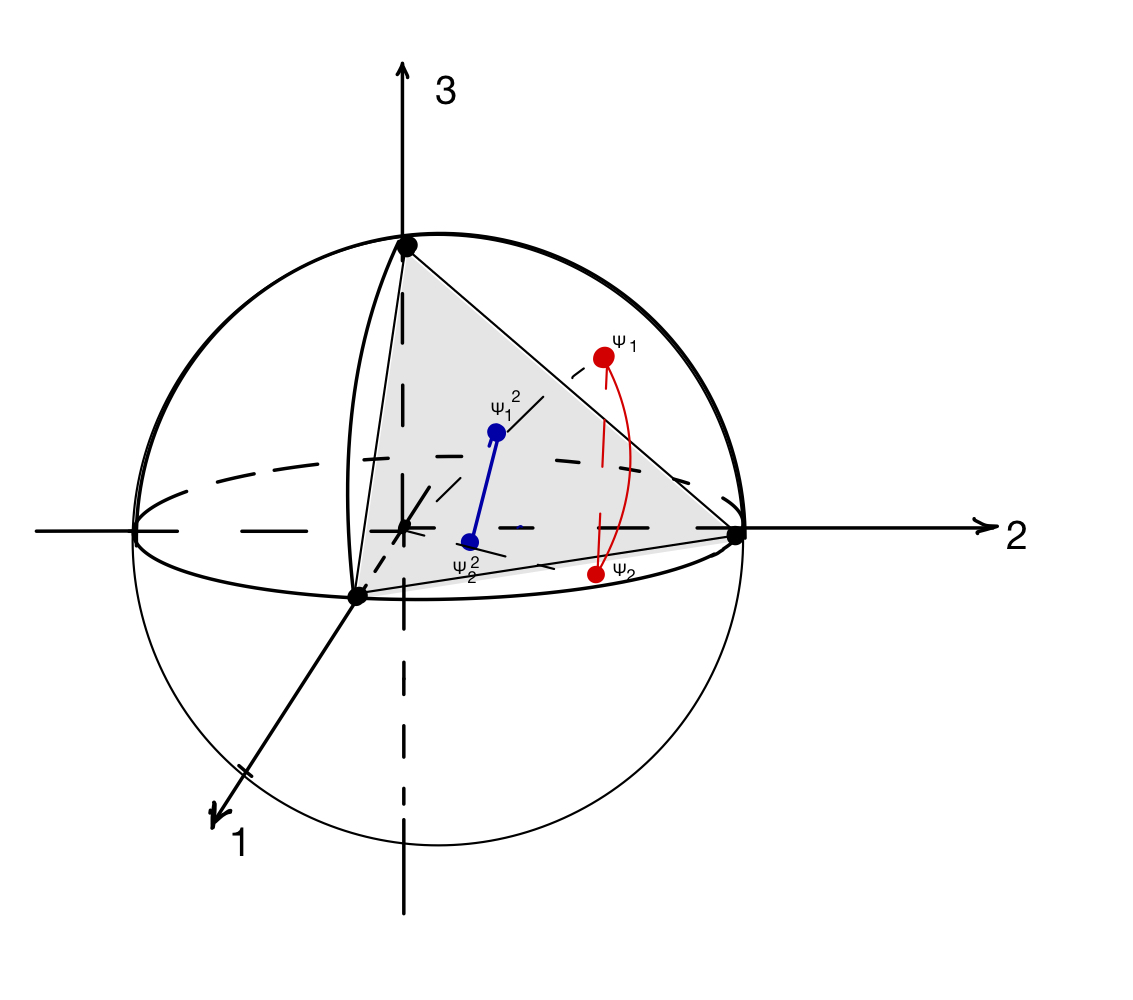} 
%		\label{fig:simple:discrete}
%		\caption{A representation of  mixture of positive weights for two components.  The blue line represents $\alpha_1\psi_1^2+\alpha_2 \psi_2^2$ on the simplex, red dashed line represents the chord $\alpha_1\psi_1+\alpha_2 \psi_2$ and its projection in $\mathcal{X}$ is the continuous red line. }		
%		\label{fig:simple:mixt}
	\end{minipage}
\caption{ (Left) A representation of  the space of discrete measures of three outcomes. The component $\psi$ in red is a point on $S^2$, the sphere of dimension two. $\psi^2$ in blue is a point in the shaded simplex whose coordinates are squares of those of $\psi$,  so that the sum of coordinates equals  one. The red cross represents some reflection of $\psi$ outside the positive orthant. (Right) A  mixture representation of positive weights for two components.  The blue line represents $\alpha_1\psi_1^2+\alpha_2 \psi_2^2$ on the simplex, red dashed line represents the chord $\alpha_1\psi_1+\alpha_2 \psi_2$ and its projection in $\mathcal{X}$ is the continuous geodesic red line. }
\label{fig:simple:discrete}
\end{figure}

Intuitively, the space of densities $Dens$ is seen as an extension as $p\to \infty$ from the discrete distributions on $p$ possible values.
See for example Figure~\ref{fig:simple:discrete} for the space of discrete measures with $p=3$ components, $x=1,2,3$, $\psi$ has only three  entries: $\psi(1)=y_1,\psi(2)=y_2,\psi(3)=y_3$ representing the red dot in the  sphere  embedded in $I\!\!R^3$ as in Figure~\ref{fig:simple:discrete}.  The resulting coordinates of $\psi^2$  are  $(y_1^2,y_2^2,y_3^2)$ represented by the blue point in Figure~\ref{fig:simple:discrete} which is in general inside the sphere. As $y_1^2+y_2^2+y_3^2=1$ the space of possible choices of $f=\psi^2$ is the shaded simplex 
while $\psi=\sqrt{f}$ are  the points of the corresponding positive orthant. 

%\begin{figure}[!tbp]
%	\centering
%	\subfloat{\includegraphics[width=0.4\textwidth]{fig2.jpeg}\label{fig:f1}}
%		\label{fig:simple:discrete}
%	\hfill
%	\subfloat{\includegraphics[width=0.4\textwidth]{fig3.jpeg}\label{fig:f2}}
%		\label{fig:simple:mixt}
%	\caption{ (a) A representation of  the space of discrete measures of three outcomes. The component $\psi$ in red is a point on $S^2$, the sphere of dimension two. $\psi^2$ in blue is a point in the shaded simplex whose coordinates are squares of those of $\psi$,  so that the sum of coordinates equals  one. The red cross represents some reflection of $\psi$ outside the positive orthant. (b) A representation of  mixture of positive weights for two components.  The blue line represents $\alpha_1\psi_1^2+\alpha_2 \psi_2^2$ on the simplex, red dashed line represents the chord $\alpha_1\psi_1+\alpha_2 \psi_2$ and its projection in $\mathcal{X}$ is the continuous red line. }
%		\label{fig:simple:discrete}
%\end{figure}
%

It is clear that by taking $p\to \infty$ and restricting the mapping $f (x)\rightarrow \psi(x)=\sqrt{f(x)}$ to the positive orthant ensures it
%\begin{eqnarray*}
%	\sqrt{\quad }&:& Dens  \rightarrow \mathcal{X}\\
%	&& f (x)\rightarrow \psi(x)=\sqrt{f(x)}
%\end{eqnarray*}
is one-to-one mapping. Therefore, one can perform analysis in the positive orthant $\mathcal{X}$ of $S^{\infty}$ and then project back onto  $Dens$. 
In particular,  due to this geometrical and analytical convenience and its connection with the Fisher-Rao metric, \cite{KB15} consider many problems of Bayesian  analysis. In particular, the $\epsilon$-contamination of the base prior for sensitivity analysis or identification of the  most influential observation are defined based on the arguments of the optimal distance between the densities within some neighbourhood of interest in $\mathcal{X}$. More recently based on the same geometric model, \cite{SBK20} consider a variational Bayesian approach. 

In this paper, we focus our attention on the optimal density choice based on minimum information and more importantly  on the neighbourhoods of interest that are defined as spanned by some class of possible prior choices  in $\mathcal{X}$.  Such densities which are considered as proposed expert opinions, will be treated as a set of linear constraints which our optimal solution is to be defined.
We will be primarily concerned with the space of differentiable densities  $
%\mathcal{X}=\{\psi|\psi(x)=\sqrt{f(x)}, x\in I \!\!R \text{ and  } 
\int \psi'^2(x) dx  <\infty .$
Our approach is based on the following important observations:
\begin{itemize}
	\item For two square-root densities in $\mathcal{X}$: $\psi_1$ and $\psi_2$, the corresponding ordinary mixture in  $Dens$ will be represented  by points $\alpha_1\psi_1^2+\alpha_2 \psi_2^2$, which for all possible weights of $ \alpha\ge 0$ parametrise  the line connecting  $\psi^2_1$ and $\psi^2_2$ (see blue segment in Figure~\ref{fig:simple:discrete}).  
	Similar expression for the counterparts in $\mathcal{X}$ as $\alpha_1\psi_1+\alpha_2 \psi_2$ for all possible values of $\alpha_i\ge 0$ will parametrise the connecting chord in the corresponding space. The radial projection of such a chord to the sphere generates the connecting geodesic  between $\psi_1$ and $\psi_2$ (see red line in Figure~\ref{fig:simple:discrete}).
	\item Unconstrained weights $\alpha_i$ including possible negative values for $\alpha_i$   in $\alpha_1\psi_1^2+\alpha_2 \psi_2^2$ will represent the plane spanned by vectors $\psi_1^2$ and $\psi_2^2$. This will present an unbounded subspace (line in $m=2$ case) beyond that of the shaded simplex $Dens$. However, the intersection  with the sphere of the corresponding plane $\alpha_1\psi_1+\alpha_2 \psi_2$  spanned by $\psi_1$ and $\psi_2$, will be simply a geodesic curve since $m=2$ while still interpretable as a density in $S^\infty$.
	
	\item In the case of  $m$  components with unconstrained weights,  the corresponding  linear combinations  $\sum_{i=1}^{m}\alpha_i\psi_i$ determine a subspace of dimension $m$. 
	In the case of densities the corresponding space  is  the projected space of square-root densities $\sum_{i=1}^{m}\alpha_i\psi_i$ onto $S^{\infty}$ and will always define a subspace of  dimension up to $m$ there. These linear combinations offer  larger and more parametrically efficient neighbourhoods of exploration than the ordinary mixtures of positive weights. 
	
	\item Once we obtain these subspaces of intersections of hyperplanes with the Hilbert sphere,  a many-to-one map from $S^{\infty} \rightarrow Dens$ can be easily constructed. Namely, that any function $\psi \in S^{\infty}$ (including those such  $\psi(x)\le 0$ from some $x$) can be mapped to some density $f$ such that $f(x)=\psi^2(x)$.
	In the discrete case as in Figure~\eqref{fig:simple:discrete} that is equivalent to assuming that $\psi$ is some point with negative coordinates like the red cross there and hence  located outside the positive orthant. However, by ignoring the coordinates sign, namely by working with absolute values $|\psi(x)|$ one can identify with that  a point in the positive orthant.  
\end{itemize}
To summarise, by working with the projected linear combinations $\sum_{i=1}^{m}\alpha_i\psi_i \in S^{\infty} $, and $|\sum_{i=1}^{m}\alpha_i\psi_i|\in \mathcal{X}$, we can explore the subspaces of dimension $m$ in $\mathcal{X}$. Additionally, as we show below, working with these subspaces enables an efficient search for opinion pooling.  

The pooling of opinions is a big area of research and has been for a number of decades. The idea is to obtain a single belief probability distribution from a set of expert opinion belief distributions. 
That is, if $(f_1,\ldots,f_n)$ represent the expert distributions,
then the aim is to  find a $f=T(f_1,\ldots,f_n)$ to represent a pooled distribution. The most common approach is linear pooling; i.e. for some weights $(\alpha_1,\ldots,\alpha_n)$, 
$f=\sum_{i=1}^n \alpha_i\,f_i.$
Other types of pooling are possible; such as log linear pooling
$\log f=K+\sum_{i=1}^n \alpha_i f_i$
which is equivalent to geometric pooling,
$f\propto \prod_{i=1}^n f_i^{\alpha_i}.$
While it would be a good constraint to impose, the $(\alpha_i)$ do not necessarily need to be weights in this case.
In principle, there is no reason why any form of pooling can not be used. In this paper, based on the motivation provided by \cite{KB15}, we work with square root pooling; i.e.
$$\sqrt{f}\propto \sum_{i=1}^n \alpha_i\,\sqrt{f_i}.$$
One clear advantage for this in addition to the motivation provided is that it is functionally and parametrically more general than linear pooling; since
$$f\propto \sum_{i=1}^n \alpha_i\,f_i+\sum_{i\ne j}\alpha_i\alpha_j\,f_i\,f_j .$$
A further reason for square--root pooling is to do with the setting of the $(\alpha_i)$ using ideas related to information. Note that we only constraint $f$ to be a density with $\sqrt{f}\in{\cal X}$, and so there is no sense in which the $(\alpha_i)$ are to be interpreted as weights in the usual sense. Indeed, they can be negative.

Our idea for the square root pooling is to derive the $(\alpha_i)$ which minimize the Fisher information for $f$. For $\psi=\sqrt{f}$ then the Fisher information is given simply  by $4 \int (\psi'(x))^2dx$. It is not possible to derive a non--trivial solution to this problem using linear pooling due to the convexity of the Fisher information; recall $I(f)=\int (f')^2/f$.  
Therefore,
$$I(f)=I(w\,f_1+(1-w)\,f_2)\leq w\,I(f_1)+(1-w)\,I(f_2),$$
and so we can obtain the minimum possible value by taking $w=0$ or 1, depending on whether $I(f_2)<I(f_1)$ or $I(f_1)<I(f_2)$, respectively. As we shall go on to show, there will be a non--trivial solution using the square root pooling provided we have all distinct $(f_i)$.

The principle we adopt for the selection of the $(\alpha_i)$ is closely related to the foundations of Maximum Entropy prior distributions. The idea is based on Jaynes (1957). 
Here entropy is the differential Shannon entropy;
$E(p)=-\int p\log p.$
This is the reverse of information; i.e. $I(p)=\int p\log p$.
Suppose the aim is to construct a prior distribution $p(x)$ which is known to satisfy a set of linear constraints,
$$\int g_j(x)\,p(x)\,d x=\mu_j,\quad \mbox{for}\quad j=1,\ldots,M.$$
One of these constraints, for $j=1$, say, involves $g_1=1$ and $\mu_1=1$. The aim then is to minimize the information subject to the constraints and this involves minimizing
$$\int p(x)\,\log p(x)\, dx+\sum_{j=1}^N \lambda_j\int g_j(x)\,p(x)\, dx,$$
where the $(\lambda_j)$ are the Lagrange multipliers. The solution is based on rewriting this expression as
$$\int p(x)\log\left\{\frac{p(x)}{\exp\left(\sum_{j=1}^M \lambda_j\,g_j(x)\right)}\right\}\,dx.$$
Based on the non--negativity of a 
Kullback--Leibler divergence, the solution is
$$p(x)\propto \exp\left(\sum_{j=1}^M \lambda_j\,g_j(x)\right),$$
where the $(\lambda_j)$ are now found to satisfy the constraints.
The following principle is the basis of the pooling; the maximum entropy (minimum information) principle states that given constraints on a prior, (it is an average of expert opinion priors)  the prior should be chosen with the maximum entropy (minimum information). Our reinterpretation of the principle is given in the brackets.

\section{Optimal prior from a finite dimensional functional space}

\begin{figure}
	\centering
	\includegraphics[scale=.07]{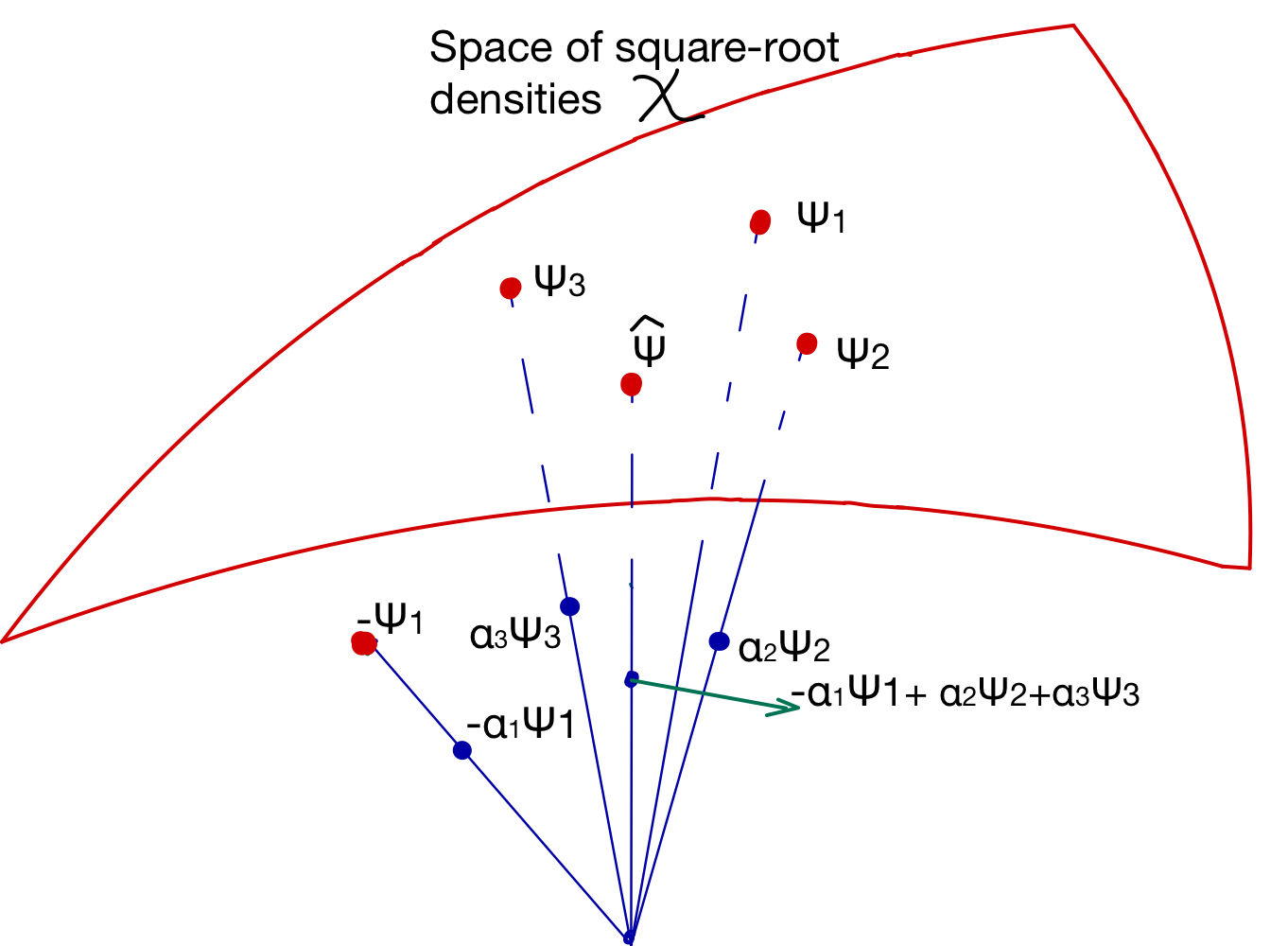} 
	\caption{A representation of  three elements and their weighted sum in the space of square-root densities. The first component has negative weight and so both $ -\psi_1$ and $-\alpha_1 \psi_1$ are outside the positive orthant but the rescaled sum $-\alpha_1 \psi_1+\alpha_2 \psi_2+\alpha_3 \psi_3$ is projected back as $\psi$. }
	\label{fig:simple:gen}
\end{figure}

Let assume that  $\boldsymbol{\psi}=(\psi_1,\psi_2,\cdots,\psi_m)^\top$ is a vector of $m$ possible prior densities in $\mathcal{X}$, and denote with $\A$ and $\B$ the $m\times m$ matrices with entries as $\A_{i,j}=< \psi_i' ,\psi_j'>=\int \psi_i'(x) \psi_j'^\top(x) d(x)$, and $\B_{i,j}=<\psi_i ,\psi_j>=\int \psi_i(x) \psi_j^\top(x) d(x)$. 
We are seeking the pooled prior as
\begin{equation}
	\hat{\psi}(x)=\frac{\sum_{i=1}^m \alpha_i \psi_i(x)}{\sqrt{\sum_{i=1}^m \alpha_i \psi_i \sum_{i=1}^m \alpha_j \psi_j}}=\frac{\pmb{\alpha}^\top}{(\pmb{\alpha}^\top\int \boldsymbol{\psi}\boldsymbol{\psi}^\top \pmb{\alpha})^{1/2} } \boldsymbol{\psi}=\frac{\pmb{\alpha}^\top}{(\pmb{\alpha}^\top \B \pmb{\alpha})^{1/2} } \boldsymbol{\psi},
%	=\frac{W^\top\psi}{\lVert W B^{1/2}\rVert }
	\label{eqn:dens:mixt}
\end{equation}
here $\alpha_i$ do not have to be positive. The numerator term ensures that the chosen mean prior is projected in the sphere $S^{\infty}$ i.e. $\int \hat{\psi}^2(x) dx=1$ and that is equivalent to assuming that $\pmb{\alpha}^\top \B \pmb{\alpha}=1$ .  Note that possible negative weights can still be producing valid density functions as $f=\psi^2$. This assumption is not too restrictive since we are searching along an infinite range of possible densities even though the subspace has dimension $m$ in the Hilbert space of  densities. One can also work with additional constraints on $\alpha_i$, such as $\alpha_i\ge0$, or even condition them such that to explore only some neighbourhood spreading along some ``contaminating directions'', as in \cite{KB15}. These could give rise to different optimisation procedures, but in essence they will provide an optimal solution on a smaller subspace than the one with unrestricted $\alpha_i$. 
It is easy to see that the corresponding information for density representation \eqref{eqn:dens:mixt} is  
\begin{equation}
4\frac{\pmb{\alpha}^\top \A \pmb{\alpha}}{\pmb{\alpha}^\top \B \pmb{\alpha}}. %\quad \text{such that , I am not sure !!! }\quad W\ge 0 
\label{eqn:Inf:mixt}
\end{equation}
This implies that, for any scalar choice $\kappa\ne0 $, the information for $\pmb{\alpha}$ and $\kappa \pmb{\alpha}$ remains unchanged.  In fact this invariance holds because  we are in effect looking for a direction in the Hilbert space as shown in Figure~\ref{fig:simple:gen}. 
Note that the matrix $\B$ resembles that of a correlation matrix as it has all its diagonal entries equal to one. 
From their construction, $ \A $ and $ \B $ are positive semidefinite because for any vector $\pmb{\alpha}$ (see e.g 7.0.2~in~\cite{HJ12} )
\[
\pmb{\alpha}^\top \A \pmb{\alpha}= \int \left( \sum_{i=1}^m \alpha_i \psi'_i(x) \right)^2 dx\ge0,  \quad \pmb{\alpha}^\top \B \pmb{\alpha}= \int \left( \sum_{i=1}^m \alpha_i \psi_i(x) \right)^2 dx\ge0,
\]
These imply that any linear combination \eqref{eqn:dens:mixt} generates a valid density as long as $\pmb{\alpha}^\top B \pmb{\alpha}>0$. 

 \begin{theorem}
	If the matrix $\B$ is not singular, the optimal $\pmb{\alpha}$ determining the density \eqref{eqn:dens:mixt} with minimal information \eqref{eqn:Inf:mixt} is any  multiple of $\B^{-1/2} \V_1$, where  $\V_1$ is the eigenvector corresponding to  the lowest eigenvalue of $\B^{-1/2} \A \B^{-1/2}$. Moreover the value of minimal information is equal \textbf{ to four times }that of the minimal eigenvalue of $\B^{-1/2} \A \B^{-1/2}$.
\end{theorem}

\textit{Proof of Theorem 1:}
As $\B$ is a positive definite matrix its square root is well defined and  so is $\Z=\B^{1/2}\pmb{\alpha}$. Therefore  the information \eqref{eqn:Inf:mixt} of $\psi$ can be equivalently defined as
$$
\frac{\Z^\top\B^{-1/2}\A \B^{-1/2} \Z}{\Z^\top \Z}=\V^\top \B^{-1/2} \A\B^{-1/2} \V,  \quad %\text{such that }\quad W=ZB^{-1/2} \ge 0 
$$
where $\V=\Z/|\Z|$ has norm one.
% and 
%\[
%\frac{\Z^\top \B^{-1/2} \A \B^{-1/2} \Z}{\Z^\top \Z}=\V^\top \B^{-1/2} \A\B^{-1/2} \V, 
%\]
The optimal vector minimising above is that unit vector corresponding to the eigenvector $\V_1$ of the smallest eigenvalue of  $\B^{-1/2} \A \B^{-1/2}$.
Therefore any multiple of $\pmb{\alpha}=\B^{-1/2} \V_1$ is a valid solution. In particular, $\pmb{\alpha}=\B^{-1/2} \V_1$ ensures that $\hat{\psi}=\pmb{\alpha}^\top \boldsymbol{\psi}$ is already in $S^\infty$ since $$\int \hat{\psi}^2(x) dx=\pmb{\alpha}^\top\B\pmb{\alpha}=\V_1^\top \V_1=1.$$ 
\endproof 

Note that the entries for matrices $\A$ and $\B$ in general can be evaluated numerically as in \cite{KB15}. However,  it is possible to avoid the numerical integration and explicitly obtain the relevant entries if the prior distributions are from the common candidates such as beta,  normal,  gamma densities (see the Supplementary Material).

\subsection{The case of $\B$ with  reduced rank}

 If the matrix $\B$ is singular, then there exists a vector of weights $\pmb{\alpha}$ such that $\pmb{\alpha}^\top \B \pmb{\alpha}=0$. Such a pathological case $\pmb{\alpha}$, could raise when there is some linear dependence among the $\psi_i$ terms. In fact these cases are not unusual as  coincident opinions among experts can occur in practise. For example, if $m=2$ and $\psi_1=\psi_2$, $\pmb{\alpha}^\top \B \pmb{\alpha}=0 $ for weights set as $-1$ and $1$ respectively.  In this case, the density is a degenerate one and the ratio \eqref{eqn:Inf:mixt}  will be  of the form $\frac{0}{0}$. In fact one can show that the null spaces of matrices $\A$ and $\B$ are related as
$
Null\B) \subseteq Null(\A)
$. 
We show this as follows. Let $\V_1,\ldots, \V_{dim(Null(\B))}$ be the 
orthogonal unit vectors spanning the null space of $\B$. Namely,
$
\V_j^\top  \V_i=\left\{ \begin{array}{cc}
1&i=j\\
0 & i\ne j
\end{array}\right. 
%\quad \text{and}\quad
%V_j^\top  B V_j=0 \quad j=1,2,\cdots, dim(Null(B) )
$
and $$\V_j^\top \B \V_j= \int \left( \sum_{i=1}^m \V_{j,i} \psi_i(x) \right)^2 dx=0,\quad j=1,2,\cdots, dim(Null(\B) ).$$ Let us focus to a particular element $\V_j$ and we will show that this is also contained in $Null(\A)$.  Since the norm of $(\sum_{i=1}^m \V_{j,i} \psi_i(x))^2$ is zero, up to a zero measurable set so is the corresponding square-root-density, i.e.
$
\sum_{i=1}^m \V_{j,i} \psi_i(x) =0.
$
Taking the differentiation inside the summation sign we then have: 
\[
\sum_{i=1}^m \V_{j,i} \psi'_i(x) =0,  \quad \text{or} \quad \int \left( \sum_{i=1}^m \V_{j,i} \psi'_i(x) \right)^2 dx=\V_{j}^\top \A \V_j=0,
\]
and so $\V_j \in Null(\A)$. 
Since these basis vectors $\V_1,\ldots, \V_{dim(Null(B)}$  are assumed orthogonal while  contained also in $Null(\A)$ it follows that  $ dim(Null(\B) )\le dim(Null(\A) ) $. 

This has an important implication since the optimal density can be easily evaluated in these cases.
Let $\B=\O \Delta \O^T$ be the eigen--decomposition and depending on the dimensionality of its null space $\Delta=\mbox{diag}(\delta_1,\ldots,\delta_r, 0,..0)$ where the number of zeros is the same as  $ dim(Null(\B) )$, and the last $ dim(Null(\B) )$ corresponding vectors of  $\O$ are the orthogonal unit vectors $\V_1,\ldots, \V_{dim(Null(\B)}$ spanning $Null(\B)$ as above. It now follows that these vectors are also contained in $Null(\A)$. Therefore a similar decomposition for $\A=\U \nabla \U^T$, $\nabla=\mbox{diag}(\xi_1,\xi_2,\ldots,\xi_r,0,\ldots,0)$ leads to the corresponding orthogonal matrix $\U$ sharing the last $m-r$ columns with those of $\O$.  Let denote the partition along these $r$ and $p-r$ vectors such that $\O=(\O_1,\O_0) $ and $ \U=(\U_1,\U_0) $ with $ \O_0=\U_0 $. Note also that 
\[
%It now follows that
%$$
\frac{\pmb{\alpha}^\top \A \pmb{\alpha}}{\pmb{\alpha}^\top \B \pmb{\alpha}}=\frac{\pmb{\alpha}^\top \U \nabla \U^T \pmb{\alpha}}{\pmb{\alpha}^\top \O \Delta \O^T \pmb{\alpha}}=\frac{\tilde{\pmb{\alpha}}^\top \R \nabla \R^\top \tilde{\pmb{\alpha}}}{\tilde{\pmb{\alpha}}^\top\Delta \tilde{\pmb{\alpha}}} ,
\]
where $\R=\O^\top \U=\left(\  \begin{array}{cc}
	\O_1^\top \U_1 &0\\
	0&I
\end{array}\right)$, and $\tilde{\pmb{\alpha}}=\O^T \pmb{\alpha}$. Let $\tilde{\pmb{\alpha}}^\top = \pmb{\alpha}^\top \O=\pmb{\alpha}^\top (\O_1,\O_0)= (\tilde{\pmb{\alpha}}^\top_1,\tilde{\pmb{\alpha}}^\top_0)$, and so
$$
\frac{\pmb{\alpha}^\top \A \pmb{\alpha}}{\pmb{\alpha}^\top \B \pmb{\alpha}}
=\frac{\tilde{\pmb{\alpha}}^\top \R \nabla \R^T\tilde{\pmb{\alpha}}}{\tilde{\pmb{\alpha}}^\top\Delta \tilde{\pmb{\alpha}}} =
\frac{\tilde{\pmb{\alpha}_1}^\top \O_1^\top \U_1 \nabla_1  \U_1^\top \O_1 \tilde{\pmb{\alpha}_1}}{\tilde{\pmb{\alpha}_1}^\top\Delta_1 \tilde{\pmb{\alpha}_1}},
$$
where $\Delta_1 =\mbox{diag}(\delta_1,\ldots,\delta_r)$ and $\nabla_1=\mbox{diag}(\xi_1,\xi_2,\ldots,\xi_r)$.
Therefore the problem is reduced as in Theorem 1, where the matrix $\B$ corresponds  now  to the non singular part  $\Delta_1$  but with dimension $r<m$ and matrix $\A$ as $\O_1^\top \U_1 \nabla_1  \U_1^\top \O_1$. The optimal information value is achieved at the smallest eigenvalue of  $  \Delta_1^{-\frac{1}{2}}  \O_1^\top \U_1 \nabla_1  \U_1^\top \O_1 \Delta_1^{-\frac{1}{2}}$.

The flexibility in our parameterization enables us to use other, alternative  basis density functions for the same problem. 
For example, if we are to replace $\psi_i$ with $\tilde{\psi}_1=\psi_1-\psi_2$, then the corresponding matrix $\tilde{\B}$ for the basis functions $\tilde{\psi}_1, \psi_2,\ldots,\psi_m$
will be
\[
%\left(\tilde{\alpha_1}\tilde{\psi}_1 +\tilde{\alpha_2}\psi_2+..+\tilde{\alpha_m}\psi_m \right)^2=W^T R BR^T W\quad 
\tilde{\B}=\R \B \R^T \quad \R=
\left(\begin{matrix}
		1&-1&0&\cdots&0\\
		0&&   &&\\
	\vdots&&\I_{m-1}&&\\
		0&&&&
	\end{matrix}\right).
%\left(\begin{matrix}
%	1&-1&0&\cdots&0\\
%	0&1&0&\cdots&0\\
%	\vdots&\vdots&\vdots&\vdots&\vdots\\
%	0&0&0&\cdots&1
%\end{matrix}\right),
\]
If the opinions say $\psi_1$  and $\psi_2$ are now coincident or getting close to each other then the corresponding matrix $\tilde{\B}$ will be having the first row and column equal (or close ) to zero. By ignoring this row and column of zeros,  the remaining block matrix $\tilde{\B}_{-1,-1}$ 
%will be the same as that of $\B$ without both the first row and first column and so 
will be of rank $m-1$.

\section{Illustrations}\label{sc_illustration}
To illustrate the proposed approach we consider initially some toy examples to show that the method proposes sensible results and outperforms the ordinary mixture pooling. We then apply our method to some real data application form the ecology where experts' opinions are combined as in \cite{Wongnak2022}. 
 
\subsection{Numerical examples for Normal densities}

\begin{figure}[ht]
 \centering
 	\begin{minipage}[b]{0.4\linewidth}
 		\centering
 		%\begin{figure}
 		\includegraphics[scale=.2]{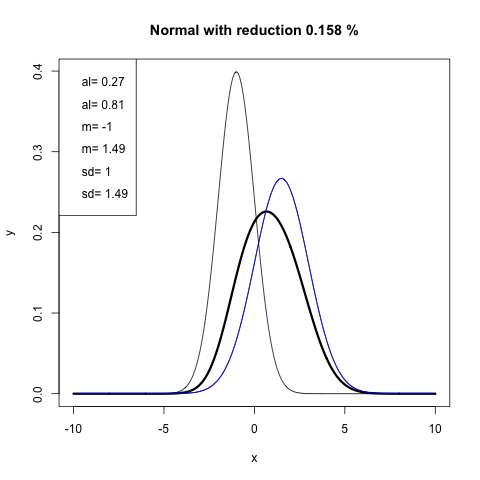} 
 		%	\caption{A representation of  the space of discrete measures of three outcomes. The component $\psi$ in red is a point on $S^2$, the sphere of dimension two. $\psi^2$ in blue is a point in the shaded simplex whose coordinates are squares of those of $\psi$,  so that the sum of coordinates equals  one. The red cross represents some reflection of $\psi$ outside the positive orthant.}
 		%\end{figure}
 	\end{minipage}
 	\hspace{0.0cm}
 	\begin{minipage}[b]{0.45\linewidth}
 		\centering
 	\includegraphics[scale=.15]{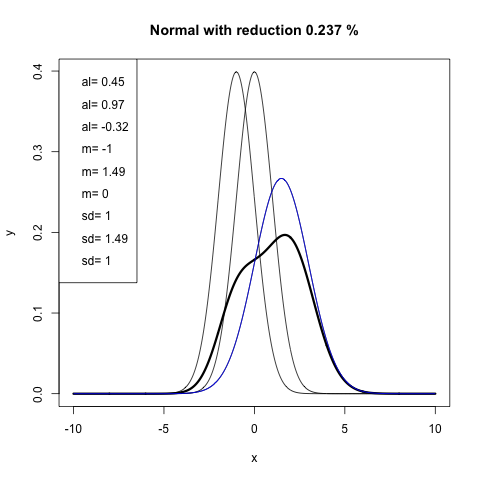} 
 		%		\label{fig:simple:discrete}
 		%		\caption{A representation of  mixture of positive weights for two components.  The blue line represents $\alpha_1\psi_1^2+\alpha_2 \psi_2^2$ on the simplex, red dashed line represents the chord $\alpha_1\psi_1+\alpha_2 \psi_2$ and its projection in $\mathcal{X}$ is the continuous red line. }		
 		%		\label{fig:simple:mixt}
 	\end{minipage}
\caption{Left: Two components case for normal distributions. The corresponding weights, mean and standard deviations are denoted at the top left of each figure respectively. Blue line represents the dominant (least informative) component and the dashed black line represents the optimal prior choice. The title represents the reduction of information from the dominant component. Right: The same as above but with three components.}
	\label{fig:toy:normal}
\end{figure}

In the first example, we consider the case of two normal densities with different means $\mu_1=-1$ and $\mu_2=1.49$ and standard errors equal to one and $1.49$ respectively. The optimal $\alpha$ values are both positive and equal to $0.27$ and $0.81$ respectively. Additionally the optimal information is reduced by $15.8\%$ compared to that of the largest variance component which the ordinary mixture solution would have produced. Similar analysis is performed for a three component case (see the right plot of Figure~\ref{fig:toy:normal} for more details about the corresponding parameters and optimal values of $\alpha$) and the information is reduced by $23.7\%$. In our extensive study for normal cases, we have noticed that the optimal pooled prior always has non zero values of $\alpha$ and its largest value is assigned to the dominant density of the largest variance, while always improving  on its information, variance in this case. 

%In the second example We report these two examples in Figure~\eqref{fig:toy:normal}. As we increase  the variance of the second component and change also its mean, the optimal mixture in our method is following this dominant component while always improving  on its information ,variance in this case. Note that in the ordinary mixture case the optimal mixture is the degenerate case at the dominant component.  See  which confirms 
% \animategraphics[controls=play,autoplay,loop, scale=.3]{12}{}{1}{100}

Similar plots are also obtained for the beta distributions (See Figure Figure~\ref{fig:toy:beta} of %section~\ref{ssec:beta:SM} 
Appendix III in the Supplementary Material%~\eqref{SM}
). 
 
\subsection{Example from Ecology}
To illustrate the proposed approach to combine experts' opinion, we consider the example in \cite{Wongnak2022}. The authors study the survival (in days) of \emph{Ixodies ricins} ticks under different (controlled) conditions of temperature and humidity. The data set analysed is available in \cite{Milne1950} (Table 16).

As Parametric Survival Model (PSM), the authors chose a Weibull distribution, where they have included in the scale parameter the effects of two covariates, temperature and humidity. That is, for tick $i\in\{1,\ldots,5\}$ and condition $j\in\{1,\ldots,20\}$, the survival time $T_{ij}$ behaves as follows:
\begin{equation}\label{eq:survivalmodel}
    T_{ij}\sim \mbox{Wei}\big(\exp(\beta_0+\beta_1 U_j^k+\beta_2Q_j+\beta_3U_j^kQ_j),p\big),
\end{equation}
where $Q_j$ and $U_j$ represent the temperature and the humidity in the experimental condition $j$, respectively, $k$ is a continuous parameter describing the degree of non-linear effects of $U_j$, $\pmb\beta=(\beta_0,\beta_1,\beta_2,\beta_3)$ is the vector of coefficients and, finally, $p$ is the shape parameter of the Weibull density. In other words, the effects of the two covariates (and their interation) are linked to the scale of the Weibull density through $\log\lambda=\exp(\beta_0+\beta_1 U_j^k+\beta_2Q_j+\beta_3U_j^kQ_j)$. It is not in the scope of this work to discuss any further the model choices described in \cite{Wongnak2022}, nor to perform any inference exercise, but only to show how the opinions of the six expects can be combined using the method here proposed.

\begin{table}[!ht]
	\centering
	
	\scalebox{0.7}{
	\begin{tabular}{llllll|llllll}
		\hline
		$e$   & $c$ & $Q_{e,c}$ & $U_{e,c}$ & $\mu_{e,c}$ & $\sigma_{e,c}$ & $e$   & $c$ & $Q_{e,c}$ & $U_{e,c}$ & $\mu_{e,c}$ & $\sigma_{e,c}$ \\ \hline
		1 & 1 & 5 & 0.30 & 3.73 & 0.73 & 4 & 1 & 5 & 0.30 & 5.19 & 1.00 \\ 
		~ & 2 & 25 & 0.30 & 2.29 & 0.52 & ~ & 2 & 20 & 0.30 & 3.38 & 1.00 \\ 
		~ & 3 & 5 & 0.95 & 4.50 & 0.21 & ~ & 3 & 5 & 0.95 & 7.15 & 0.29 \\
		~ & 4 & 25 & 0.95 & 5.01 & 0.10 & ~ & 4 & 20 & 0.80 & 6.31 & 0.47 \\
		2 & 1 & 5 & 0.10 & 2.30 & 0.32 & 5 & 1 & 5 & 0.10 & 0.68 & 0.26 \\ 
		~ & 2 & 25 & 0.10 & 1.60 & 0.30 & ~ & 2 & 25 & 0.10 & 0.68 & 0.26 \\ 
		~ & 3 & 7 & 0.90 & 5.52 & 0.30 &  ~ & 3 & 5 & 0.95 & 5.90 & 0.17 \\ 
		~ & 4 & 20 & 0.90 & 4.09 & 0.52 & ~ & 4 & 25 & 0.95 & 5.60 & 0.18 \\ 
		3 & 1 & 5 & 0.10 & 2.69 & 0.81 &  6 & 1 & 5 & 0.10 & 2.05 & 0.73 \\ 
		~ & 2 & 25 & 0.10 & 1.94 & 0.37 &  ~ & 2 & 25 & 0.10 & 0.57 & 0.69 \\
		~ & 3 & 5 & 0.95 & 5.72 & 0.16 &  ~ & 3 & 8 & 0.95 & 4.79 & 0.25 \\ 
		~ & 4 & 20 & 0.95 & 5.90 & 0.05 & ~ & 4 & 25 & 0.95 & 2.89 & 0.43 \\ \hline
	\end{tabular}
}	\caption{The opinion of the experts on the average survival time.: Expert ($e$), Condition ($c$), Temperature ($Q_{e,c}$), Humidity ($U_{e,c}$) Location ($\mu_{e,c}$) and Scale ($\sigma_{e,c}$).}
	\label{tab:expertsopinion}
\end{table}

The opinions of six experts on the average survival time have been collected, processed and, for each one of them, the believes have been represented by a Log-normal density with given location and scale parameters. The process to transform prior information (i.e. opinion) into suitable density distributions is discussed in the detail in \cite{Wongnak2022} and we refer the reader to it. In short, each expert was asked to provide an informed ``guess'' of the mean survival time, the highest and the lowest survival time and, finally, how confident they were about their ``guesses''. The information, through a suitable software (and some revision activities) was then transformed in the parameters of the log normal density representing their uncertainty.
As such, for expert $e$ and experimental condition $c$, the average survival time had the following distribution:
$\overline{T}_{e,c} \sim Ln(\mu_{e,c},\sigma_{e,c}),$
where $\mu$ and $\sigma$ where the location and scale parameter, respectively, of the Log-normal density determined above.
\begin{figure}[!ht]
    \centering
    \includegraphics[scale=0.35]{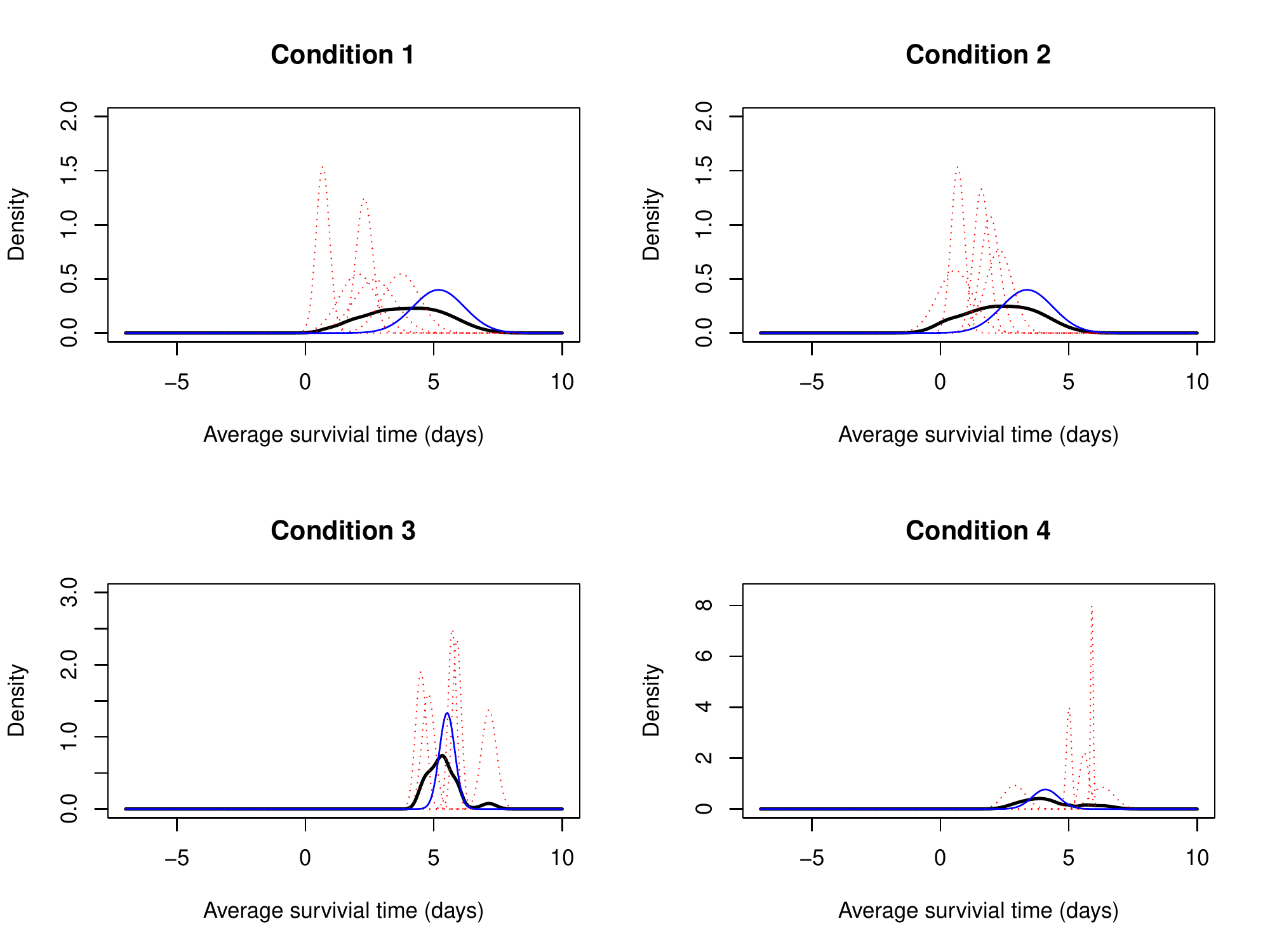}
    \caption{Combining prior experts' opinions.For each Condition, we have the normal densities representing the average survival time in the exponential scale (red dashed lines) and the combined density (thick solid black line). We have also highlighted (solid blue line) the density that would have been chosen to represent the combined expert opinions if one would consider maximizing the information in the prior.}
    \label{fig:expertslognormal}
\end{figure}

In \cite{Wongnak2022} the aggregation of the experts' opinions was performed at the level of the induced priors on $\pmb\beta$. Without questioning this approach, as our aim is to illustrate the aggregation process only, we decided to aggregate at the root of the model. In other words, as the prior information was used to formulate a prior distribution on the (mean) survival time, we have combined the information contained in the log-normal densities.
The parameters for the log-normal densities are contained in Table \ref{tab:expertsopinion}.

In Figure \ref{fig:expertslognormal} we have included, for each condition, the resulting density obtained by combining the experts' opinions, represented by normal densities. We have chosen to not represent the aggregation in the log scale so to have a better visual representation of the differences between our aggregation method and the resulting density one would select if the information was maximized. It is clear the the proposed method results in a wider representation of the experts' opinions, and the resulting combined densities (for each condition) assign more mass to the region where, overall, the curves representing the experts' opinions are more dense. This is particularly obvious for Conditions 1 and 2. Interestingly, in Condition 3, we note that the combined density is bimodal, hence capable to incorporate the opinion of the expert represented by the density to the far right. Although less prominent, the same happens for Condition 4.

\newpage
%\section{Appendix}
\section*{Supplementary material}
\setcounter{section}{4}
\setcounter{subsection}{0}
\label{SM}
%Further material such as technical details, extended proofs, code, or additional  simulations, figures and examples may appear online, and should be briefly mentioned as Supplementary Material where appropriate.  Please submit any such content as a PDF file along with your paper, entitled `Supplementary material for Title-of-paper'.  After the acknowledgements, include a section `Supplementary material' in your paper, with the sentence `Supplementary material available at \Bka\ online includes $\ldots$', giving a brief indication of what is available.  However it should be possible to read and understand the paper without reading the supplementary material.
%
%Further instructions will be given when a paper is accepted.
\section*{Appendix I}
\subsection*{$\A$ and $\B$  entries for  exponential family components} 

For the general exponential family parameterization: 
\[
f(x,\theta)=\exp\{ \eta(\theta)T(x)-A(\theta)+B(x)\},
\]
\[
\psi(x,\theta)=f(x,\theta)^{1/2}=\exp\left\{ \frac{\eta(\theta)T(x)-A(\theta)+B(x)}{2}\right\},
\quad \text{and}\quad 
\psi'(x,\theta)=\frac{1}{2}(\eta(\theta)T'(x)+B'(x))\psi(x,\theta),
\]
and for a pair $\psi_1$ and $\psi_2$ we could write
\[
<\psi_1,\psi_2>=\exp\left\{ \frac{-A_1(\theta_1)-A_2(\theta_2)}{2}\right\} \int \exp\left\{ \frac{\eta_1(\theta_1)T_1(x)+B_1(x)+\eta_2(\theta_2)T_2(x)+B_2(x)}{2}\right\}dx,
\]
and
\[
<\psi'_1,\psi'_2>=\frac{1}{4}\int \left\{ (\eta_1(\theta_1)T_1'(x)+B_1'(x))(\eta_2(\theta_2)T_2'(x)+B_2'(x)) \right \} \psi_1(x,\theta_1) \psi_2(x,\theta_2)dx.
\]
These indicate that it is possible to explicitly  obtain entries of matrices $\A$ and $\B$ where the prior distributions are from the common candidates such as beta,  normal,  gamma, inverse gamma densities. We report in the following these pairwise entries explicitly.
\section*{Appendix II}
\subsection*{Calculations for normal densities}
For any two particular normal densities 
\[
f_i(x,\mu_i, \sigma_i)=(2\pi \sigma^2_i)^{-1/2} \exp \left\{-\frac{(x-\mu_i)^2}{2 \sigma_i^2}\right\}, \quad f_j(x,\mu_j, \sigma_j)=(2\pi \sigma^2_j)^{-1/2} \exp \left\{-\frac{(x-\mu_j)^2}{2 \sigma_j^2}
\right\},\] 
\begin{eqnarray*}
	\psi_i(x)\psi_j(x)=(2\pi \sigma^2_i)^{-1/4} (2\pi \sigma^2_j)^{-1/4}\exp \left\{-\frac{(x-\mu_i)^2}{4 \sigma_i^2}-\frac{(x-\mu_i)^2}{4 \sigma_j^2}\right\},
\end{eqnarray*} 
and by completing the relevant quadratic terms 
\[
<\psi_i,\psi_j>=(2\pi \sigma^2_i)^{-1/4} (2\pi \sigma^2_j)^{-1/4} \exp \left\{\frac{\beta^2}{4\alpha}-\frac{1}{4}\left(\frac{\mu_i^2}{\sigma_i^2}+\frac{\mu_j^2}{\sigma_j^2}\right) \right\} \sqrt{\frac{2 \pi 2}{\alpha}} \underbrace{\int f(x,\mu=\beta/\alpha,\sigma=\sqrt{2/\alpha}) dx}_{=1},
\]
for $\alpha=1/\sigma_i^2+1/\sigma_j^2$ and $\beta=\mu_i/\sigma_i^2+\mu_j/\sigma_j^2$.
This leads to
\begin{eqnarray}
	<\psi_i,\psi_j>
	&=&\sqrt{\frac{\alpha}{2\sigma_i\sigma_j}}\exp \left\{\frac{\beta^2}{4\alpha}-\frac{1}{4}\left(\frac{\mu_i^2}{\sigma_i^2}+\frac{\mu_j^2}{\sigma_j^2}\right) \right\}.
\end{eqnarray}
Similar calculations for $\psi'_i$ show that
\begin{eqnarray}
	<\psi'_i ,\psi'_j>
	&=&\frac{<\psi_i,\psi_j>}{4\sigma_i^2 \sigma_j^2}\left( \frac{2}{\alpha}+\frac{\beta^2}{\alpha^2}-\frac{\beta(\mu_i+\mu_j)}{\alpha}+\mu_i\mu_j\right).
\end{eqnarray}

\section*{Appendix III}
\subsection*{Calculations for Beta densities}

\begin{figure}[ht]
	\begin{minipage}[b]{0.32\linewidth}
		\centering
		%\begin{figure}
		\includegraphics[scale=.2]{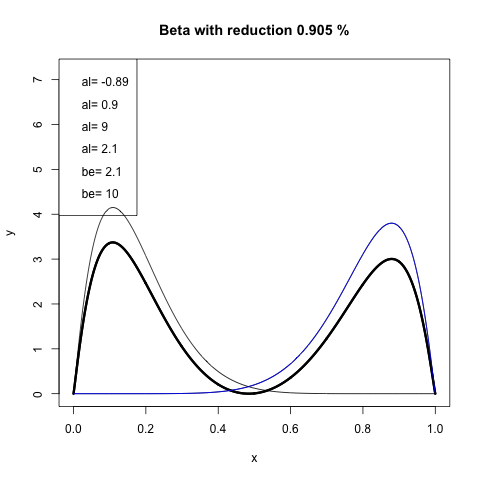} 
		%	\caption{A representation of  the space of discrete measures of three outcomes. The component $\psi$ in red is a point on $S^2$, the sphere of dimension two. $\psi^2$ in blue is a point in the shaded simplex whose coordinates are squares of those of $\psi$,  so that the sum of coordinates equals  one. The red cross represents some reflection of $\psi$ outside the positive orthant.}
		%\end{figure}
	\end{minipage}
	\hspace{0.0cm}
	\begin{minipage}[b]{0.32\linewidth}
		\centering
		\includegraphics[scale=.2]{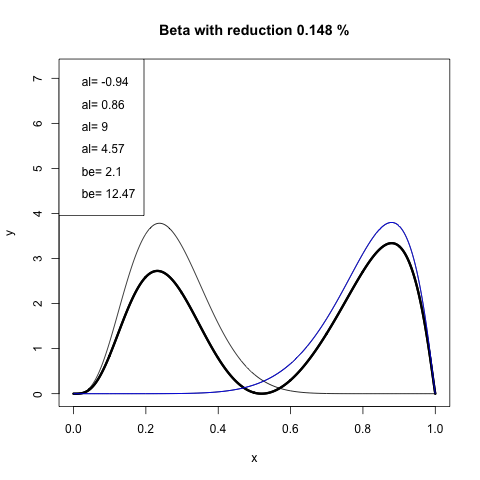} 
		%		\label{fig:simple:discrete}
		%		\caption{A representation of  mixture of positive weights for two components.  The blue line represents $\alpha_1\psi_1^2+\alpha_2 \psi_2^2$ on the simplex, red dashed line represents the chord $\alpha_1\psi_1+\alpha_2 \psi_2$ and its projection in $\mathcal{X}$ is the continuous red line. }		
		%		\label{fig:simple:mixt}
	\end{minipage}
	\begin{minipage}[b]{0.32\linewidth}
		\centering
		\includegraphics[scale=.25]{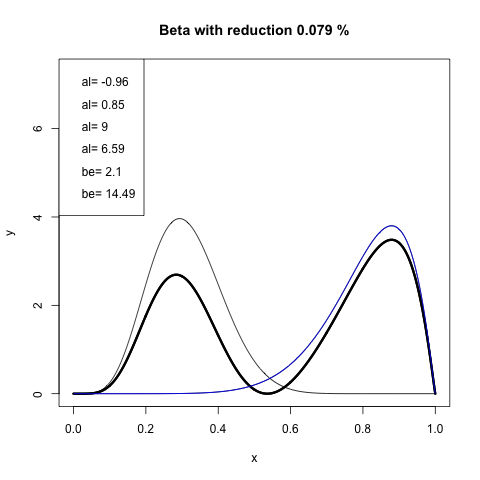} 
		%		\label{fig:simple:discrete}
		%		\caption{A representation of  mixture of positive weights for two components.  The blue line represents $\alpha_1\psi_1^2+\alpha_2 \psi_2^2$ on the simplex, red dashed line represents the chord $\alpha_1\psi_1+\alpha_2 \psi_2$ and its projection in $\mathcal{X}$ is the continuous red line. }		
		%		\label{fig:simple:mixt}
	\end{minipage}
	% 	\caption{  }
	% 	\label{fig:toy:normal}
	% \end{figure}
%
% \begin{figure}[ht]
	\begin{minipage}[b]{0.32\linewidth}
		\centering
		%\begin{figure}
		\includegraphics[scale=.25]{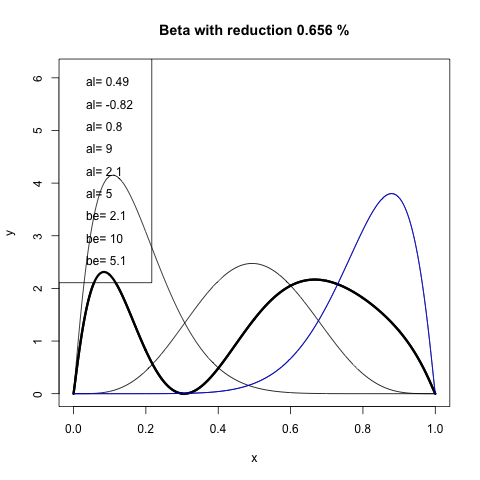} 
		%	\caption{A representation of  the space of discrete measures of three outcomes. The component $\psi$ in red is a point on $S^2$, the sphere of dimension two. $\psi^2$ in blue is a point in the shaded simplex whose coordinates are squares of those of $\psi$,  so that the sum of coordinates equals  one. The red cross represents some reflection of $\psi$ outside the positive orthant.}
		%\end{figure}
	\end{minipage}
	\hspace{0.0cm}
	\begin{minipage}[b]{0.32\linewidth}
		\centering
		\includegraphics[scale=.25]{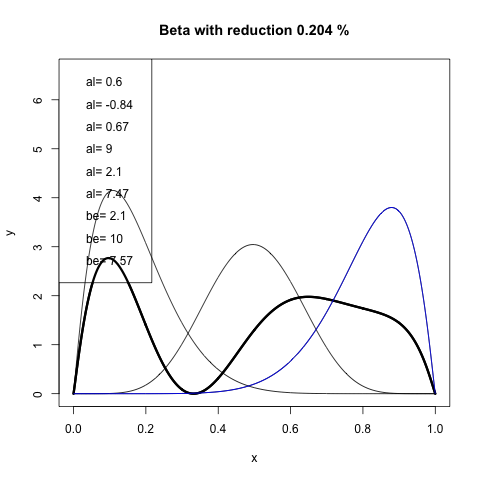} 
		%		\label{fig:simple:discrete}
		%		\caption{A representation of  mixture of positive weights for two components.  The blue line represents $\alpha_1\psi_1^2+\alpha_2 \psi_2^2$ on the simplex, red dashed line represents the chord $\alpha_1\psi_1+\alpha_2 \psi_2$ and its projection in $\mathcal{X}$ is the continuous red line. }		
		%		\label{fig:simple:mixt}
	\end{minipage}
	\begin{minipage}[b]{0.32\linewidth}
		\centering
		\includegraphics[scale=.25]{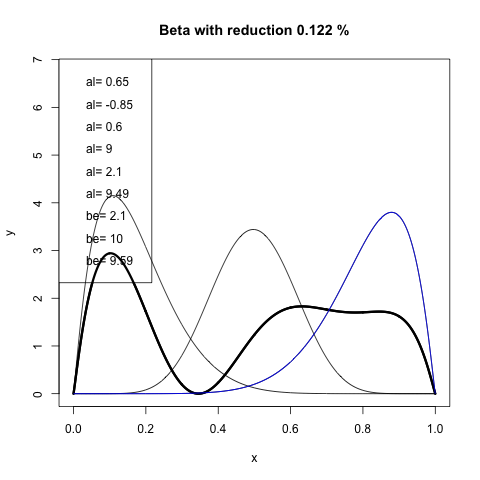} 
		%		\label{fig:simple:discrete}
		%		\caption{A representation of  mixture of positive weights for two components.  The blue line represents $\alpha_1\psi_1^2+\alpha_2 \psi_2^2$ on the simplex, red dashed line represents the chord $\alpha_1\psi_1+\alpha_2 \psi_2$ and its projection in $\mathcal{X}$ is the continuous red line. }		
		%		\label{fig:simple:mixt}
	\end{minipage}	
	\caption{Top Row: Two components case for beta distributions. The corresponding weights, mean and standard deviations are denoted at the top left of each figure respectively. Blue line represents the dominant (least informative) component, dashed black line represents  the optimal prior choice. Bottom Row: The same as above but with three components.  }
	\label{fig:toy:beta}
\end{figure}

For the $\mbox{Beta}(\alpha_i,\beta_i)$ distribution
\[
\psi_i^2=B(\alpha_i,\beta_i) x^{\alpha_i-1} (1-x)^{\beta_i-1},\quad B(\alpha_i,\beta_i)=\frac{\Gamma(\alpha_i) \Gamma(\beta_i)}{\Gamma(\alpha_i+\beta_i)}=\int_{0}^{1}  x^{\alpha_i-1} (1-x)^{\beta_i-1}dx,
\] 
and 
\begin{eqnarray*}
	\psi'_i&=&
	B(\alpha_i,\beta_i)^{-1/2} 
	\left(  
	\frac{\alpha_i-1}{2} x^{\frac{\alpha_i-3}{2}} (1-x)^{\frac{\beta_i-1}{2}}+\frac{\beta_i-1}{2} x^{\frac{\alpha_i-1}{2}} (1-x)^{\frac{\beta_i-3}{2}}	\right),	 
\end{eqnarray*}
Then it is possible to show that
\begin{eqnarray*}
	<\psi_i,\psi_j>&=&\int B(\alpha_i,\beta_i)^{-1/2} x^{\frac{\alpha_i-1}{2}} (1-x)^{\frac{\beta_i-1}{2}} B(\alpha_j,\beta_j)^{-1/2} x^{\frac{\alpha_j-1}{2}} (1-x)^{\frac{\beta_j-1}{2}} dx\\
	&=&\frac{B(\frac{\alpha_i+\alpha_j}{2},\frac{\beta_i+\beta_j}{2})}{ B(\alpha_i,\beta_i)^{1/2} B(\alpha_j,\beta_j)^{1/2}},
\end{eqnarray*}
and
\begin{eqnarray*}
	<\psi'_i,\psi'_j>&=&\int B(\alpha_i,\beta_i)^{-1/2} 
	\left(  
	\frac{\alpha_i-1}{2} x^{\frac{\alpha_i-1}{2}-1} (1-x)^{\frac{\beta_i-1}{2}}+\frac{\beta_i-1}{2} x^{\frac{\alpha_i-1}{2}} (1-x)^{\frac{\beta_i-1}{2}-1}	\right)	\\
	&\times&
	B(\alpha_j,\beta_j)^{-1/2}	\left(  
	\frac{\alpha_j-1}{2} x^{\frac{\alpha_j-1}{2}-1} (1-x)^{\frac{\beta_j-1}{2}}+\frac{\beta_j-1}{2} x^{\frac{\alpha_j-1}{2}} (1-x)^{\frac{\beta_j-1}{2}-1}	\right)	dx\\
	&=&	\frac{4^{-1}}{B(\alpha_i,\beta_i)^{1/2} B(\alpha_j,\beta_j)^{1/2}}\left(
	\frac{B(\frac{\alpha_i+\alpha_j}{2}-2,\frac{\beta_i+\beta_j}{2})}{(\alpha_i-1)^{-1}(\alpha_j-1)^{-1}}+
	\frac{B(\frac{\alpha_i+\alpha_j}{2}-1,\frac{\beta_i+\beta_j}{2}-1)}{(\alpha_i-1)^{-1}(\beta_j-1)^{-1}}\right.\\
	&+&
	\left.\frac{B(\frac{\alpha_i+\alpha_j}{2}-1,\frac{\beta_i+\beta_j}{2}-1)}{(\beta_i-1)^{-1}(\alpha_j-1)^{-1}}+
	\frac{B(\frac{\alpha_i+\alpha_j}{2},\frac{\beta_i+\beta_j}{2}-2)}{(\beta_i-1)^{-1}(\beta_j-1)^{-1}}
	\right).
\end{eqnarray*}
See Figure~\ref{fig:toy:beta} for some examples for beta densities using these calculations. 

\section*{Appendix IV}
\subsection*{Calculations for the gamma distributions} 
The parameterization of the gamma distribution as
$
\psi^2(x)=\frac{\beta^\alpha}{\Gamma(\alpha)} x^{\alpha-1}e^{-\beta x}
$
leads to
\[
\psi(x)=f(x)^{1/2}=\frac{\beta^\frac{\alpha}{2}}{\Gamma(\alpha)^{1/2}} x^{\frac{\alpha-1}{2}}e^{-\frac{\beta}{2} x},
\quad \text{and}\quad
\psi'(x)=\left(\frac{\alpha-1}{2 x}  -\frac{\beta}{2}\right)\psi(x).
\]
Therefore
\begin{eqnarray*}
	<\psi_i,\psi_j>&=&\int \frac{\beta_i^\frac{\alpha_i}{2}}{\Gamma(\alpha_i)^{1/2}} x^{\frac{\alpha_i-1}{2}}e^{-\frac{\beta_i}{2} x} \frac{\beta_j^\frac{\alpha_j}{2}}{\Gamma(\alpha_j)^{1/2}} x^{\frac{\alpha_j-1}{2}}e^{-\frac{\beta_j}{2} x}dx\\
	&=&	\frac{\beta_i^\frac{\alpha_i}{2}}{\Gamma(\alpha_i)^{1/2}} \frac{\beta_j^\frac{\alpha_j}{2}}{\Gamma(\alpha_j)^{1/2}}\frac{\Gamma(\frac{\alpha_i+\alpha_j}{2})}{(\frac{\beta_i+\beta_j}{2})^{\frac{\alpha_i+\alpha_j}{2}}},
\end{eqnarray*}
and
\begin{eqnarray*}
	<\psi'_i,\psi'_j>
	&=&\frac{\alpha_i-1}{2 } \frac{\alpha_j-1}{2 }\frac{\beta_i^\frac{\alpha_i}{2}}{\Gamma(\alpha_i)^{1/2}} \frac{\beta_j^\frac{\alpha_j}{2}}{\Gamma(\alpha_j)^{1/2}}\frac{\Gamma(\frac{\alpha_i+\alpha_j}{2}-2)}{(\frac{\beta_i+\beta_j}{2})^{\frac{\alpha_i+\alpha_j}{2}-2}}+\frac{\beta_i \beta_j}{4}\frac{\beta_i^\frac{\alpha_i}{2}}{\Gamma(\alpha_i)^{1/2}} \frac{\beta_j^\frac{\alpha_j}{2}}{\Gamma(\alpha_j)^{1/2}}\frac{\Gamma(\frac{\alpha_i+\alpha_j}{2})}{(\frac{\beta_i+\beta_j}{2})^{\frac{\alpha_i+\alpha_j}{2}}}\\
	&-&\left(\frac{\alpha_i-1}{2 }\frac{\beta_j}{2}  +\frac{\alpha_j-1}{2 } \frac{\beta_i}{2}\right)\frac{\beta_i^\frac{\alpha_i}{2}}{\Gamma(\alpha_i)^{1/2}} \frac{\beta_j^\frac{\alpha_j}{2}}{\Gamma(\alpha_j)^{1/2}}\frac{\Gamma(\frac{\alpha_i+\alpha_j}{2}-1)}{(\frac{\beta_i+\beta_j}{2})^{\frac{\alpha_i+\alpha_j}{2}-1}}.
\end{eqnarray*}
For the exponential densities with parameterization as the gamma distribution above with $\alpha=1$, then
$\psi_i(x)=\beta_i e^{-\beta_i x/2}$
and
\[
<\psi_i,\psi_j>=\frac{2 \beta_i \beta_j}{\beta_i+\beta_j},\quad\mbox{and}\quad <\psi'_i,\psi'_j>=\frac{4 \beta_i \beta_j}{(\beta_i+\beta_j)^2}.
\]
%\section*{Appendix V}
%\subsection*{Numerical illustrations for Beta distributions}
%\label{ssec:beta:SM}

\end{document}